# Design and Construction of a Dedicated Radiolucent 8-element Flexible Radiofrequency (RF) Torso Coil for the 1.0T Australian MRI-Linac System

Mingyan Li[1], Ewald Weber[1], David Waddington[2], Shanshan Shan[2], Paul Liu[2], Bin Dong[3], Paul Keall[2], Feng Liu[1] and Stuart Crozier[1]

1. School of Information Technology and Electrical Engineering, The University of Queensland, St Lucia, Queensland, Australia, 4072
2. Image X Institute, Sydney School of Health Sciences, Faculty of Medicine & Health, The University of Sydney, NSW, Australia, 2015
3. Ingham Institute for Applied Medical Research, Liverpool, NSW, Australia 2170

## Abstract

Magnetic resonance imaging-guided linear accelerators (MRI-Linacs) are an emerging treatment technology that enable online soft-tissue visualisation and adaptive radiotherapy. The Australian 1.0 T MRI-Linac employs a fixed, inline beamline, with treatment angles achieved via patient couch rotation, and currently relies on a radiolucent whole-body radiofrequency (RF) coil. However, the large coil-patient separation limits signal-to-noise ratio (SNR) and constrains the fast imaging required for real-time tumour tracking. Here, we show the design and validation of a dedicated, flexible 8-element receive-only RF torso array tailored to the geometry and beam access requirements of the Australian MRI-Linac. Electromagnetic simulations, bench measurements, and phantom imaging demonstrate approximately three-fold SNR improvement at the centre and eight-fold improvement at superficial regions compared with the whole-body coil. In vivo abdominal imaging confirms enhanced anatomical detail and sharper tissue boundaries, and the increased SNR enables parallel imaging with GRAPPA acceleration factors up to three without severe artefacts. Dosimetry measurements and dynamic beam-on noise analysis confirm that the coil array is effectively radiolucent, with dose differences below 1% and no measurable degradation in image noise. This work demonstrates a practical RF solution for accelerated, high-SNR imaging compatible with patient rotation and beam delivery on the Australian MRI-Linac.

## 1. Introduction

Magnetic Resonance Imaging (MRI) uses abundant hydrogen protons in the human body to produce images with superior soft-tissue contrast. Aligned with its versatile contrast mechanisms, MRI has become one of the preferred imaging modalities to investigate internal anatomies. Therefore, employing MRI as an image-guidance and adaptive-planning tool has become a popular approach to improve radiotherapy outcomes in MRI-guided radiotherapy (MRIgRT) over the last decades, [1,2]. Two commercial systems, the Elekta[3] and ViewRay[4] both use a perpendicular configuration, in which the main magnetic field ($B_0$) is perpendicular to the X-ray beam from the linear accelerator (Linac). During radiotherapy in these systems, the patient is in the supine position and the Linac beam is delivered from various angles by moving the Linac gantry around the patient. However, in this perpendicular configuration, the Lorentz force could deviate the electron trajectory, affect dose deposition and challenge treatment planning[5-8]. Therefore, the Australian MRI-Linac system [9] and the University of Alberta/Magnettx[10] MRI-Linacs adopt the inline setup, where the Linac position is fixed and the X-ray beam is parallel to $B_0$ to avoid such issues. On the Australian system, the large opening of the split magnet offers more versatile treatment options, such as the patient being in a standing position, sitting position, or being rotated via a rotating couch[11].

For routine imaging, the 1T Australian MRI-Linac has a whole-body radiofrequency (RF) coil. After a series of assessments, the radiolucent RF coil was proved capable of producing reasonably high-quality images at 1T[12]. However, like other whole-body coils, this transceive coil is positioned far from the patient; thus, the signal-to-noise ratio (SNR) is compromised by the long distance between the coil and the patient. One of the most efficient methods to boost the SNR is using an array of receive coils placed in close proximity to the patient[13]. In addition, the multi-channel coil array can accelerate imaging and facilitate real-time tumor tracking by enabling a simultaneous acquisition mode, which the whole-body coil does not offer. By using the multi-element receive coil array, the scan time can be easily reduced to one-third with Parallel Imaging[14,15] or even more with the combination of Compressed Sensing[16] and Deep Learning techniques[17].

Designing the phased-array coils for the MRI-Linac system is different from a standard MRI system. For example, although the Elekta MRI-Linac system (Unity, Elekta AB, Stockholm, Sweden) uses 8-element RF coils, two 4-element sub-arrays need to be positioned centimeters away from patient[18] to avoid secondary electron radiation doses[19]. A 32-element dense receive-only RF coil array with low-density foam spacers[20] was designed and partially manufactured to reduce the additional surface radiation and further boost the SNR. Although the surface SNR increases and parallel imaging ability is enhanced, the center SNR may not rise proportionally. As shown in work[21], the surface SNR of the 96-channel RF coil array dramatically increased compared to the 32 and 12-channel coils, but the center SNR of the 96-channel RF coil array is 20% lower than the 32-channel RF coil array. Zijlema[20] also demonstrated that the center SNR of the 32-channel dense array has no difference from the 8-element clinical RF coil array. Therefore, when designing the multi-channel array coils for the MRI-Linac system, the balance among all aspects needs to be carefully considered dependent to the applications.

The Australian MRI-Linac system uses a fixed beam-gantry and a rotating couch to deliver radiotherapy at different angles. The patient couch has the ability to remotely and precisely position the patient within a range ± 90 degrees from the normal 0degree supine position. During the couch rotation, the RF coil and the cables will inevitably frequently move with the patient simultaneously, therefore the array's robustness is prioritized. Considering many tumors are normally deeply located in the body, such as lung cancer, prostate cancer, liver, and cervix cancer; in this work, we chose to develop an 8-element receive-only torso coil array with a balanced consideration of penetration depth, SNR improvement, and engineering robustness. This 2 x 4 RF coil array was first numerically simulated and compared against the current whole-body coil in terms of SNR, then manufactured with flexible co-axial cables to realize multiple close-fitting schemes compatible with different treatment sites and beam angles. The image quality was evaluated at different treatment angles when the bed was rotated. The performance of the proposed coil array and the wholebody coil was compared with MR images acquired from a water phantom and volunteers. It is proved that the proposed coil array has a superior SNR advantage over the existing whole-body coil. Additionally, the fast-imaging ability of the coil array was also tested up to an acceleration factor of three. The consistent dosimetry readings with and without the presence of the torso coil array demonstrated the radiolucency of the coil.

## 2. Methods and Materials

### 2.1 Numerical simulations

The proposed RF coil array was firstly simulated to estimate its SNR improvement over the existing whole-body coil. The simulation was performed on an electromagnetic simulation platform, Sim4Life (ZMT, Zurich, Switzerland), which is based on the Finite-difference time-domain (FDTD)[22] method. The proposed 8-element RF coil array consists of two sub-arrays with 4 elements. Given the diameter of spherical volume (DSV) of the Australian MRI-Linac system is 300 mm (±4.05 ppm), the length of the individual coil element is 300 mm, and the width of each coil is 100 mm. As shown in Figure 1(a), the coils were overlapped for decoupling, which gives in total 350 mm width for each sub-array. The red dashed boxes denote the radiation windows that the beams can pass through without touching the coil wires. A water phantom and the Duke human model from the Virtual Family [23] were used for SNR analysis. The relative permittivity and conductivity of the water phantom were set to 85 and 0.3 S/m in the simulation. The Duke model and water phantom setup in the MRI-Linac whole-body coil and on the 8-element RF torso coil array are shown in Figures 1(b) and (c), respectively.

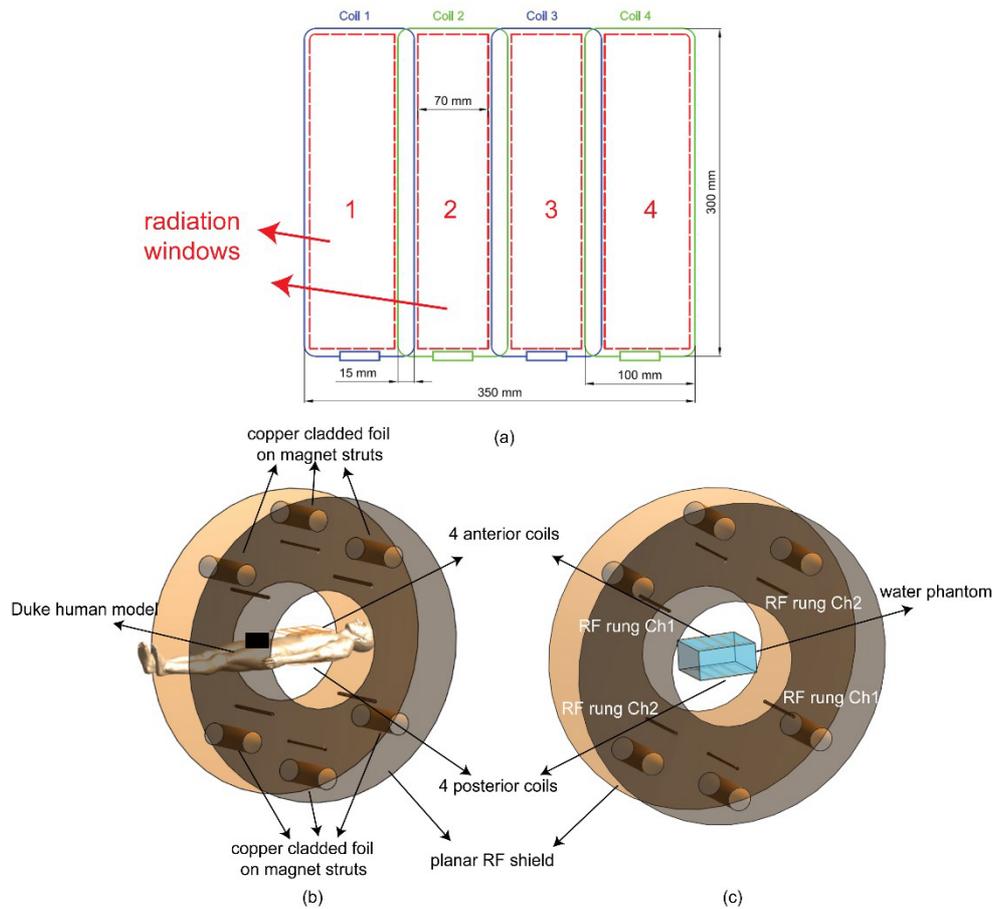

Figure 1 (a) Coil geometry and arrangement of the sub-array. (b) Simulation setup with the Duke model in the Sim4Life environment. (c) Simulation setup with the water phantom model in the Sim4Life environment for SNR calculation.

2.2 Coil manufacturing and experiment setup

To minimize interference between the coil array and the Linac beam, a design was chosen which does not need placement of any active or passive electronic components in the imaging and treatment area. It is also important to maximize the size of the anterior/posterior radiation windows and assure conformal fitting to the patient for different imaging setups. Therefore, non-magnetic thin flexible co-axial cable (Φ: 1.87mm, ENVIROFLEX-178, HUBER+SUHNER AG, Switzerland) was used to build the eight individual coil elements. The distributed capacitance formed between the woven copper shield and the core copper conductor allows tuning to an inherent resonant frequency close to 42 MHz without adding capacitors in the loop coils. These advantages make the coaxial cable the optimal material to manufacture this torso coil. This torso coil array was fitted on a volunteer and a water phantom as shown in Figures 2(b) and 3(a). The large open bore of the system also allows the patient to be treated in various positions, such as the standing position, as shown in Figures 2(c) and (d).

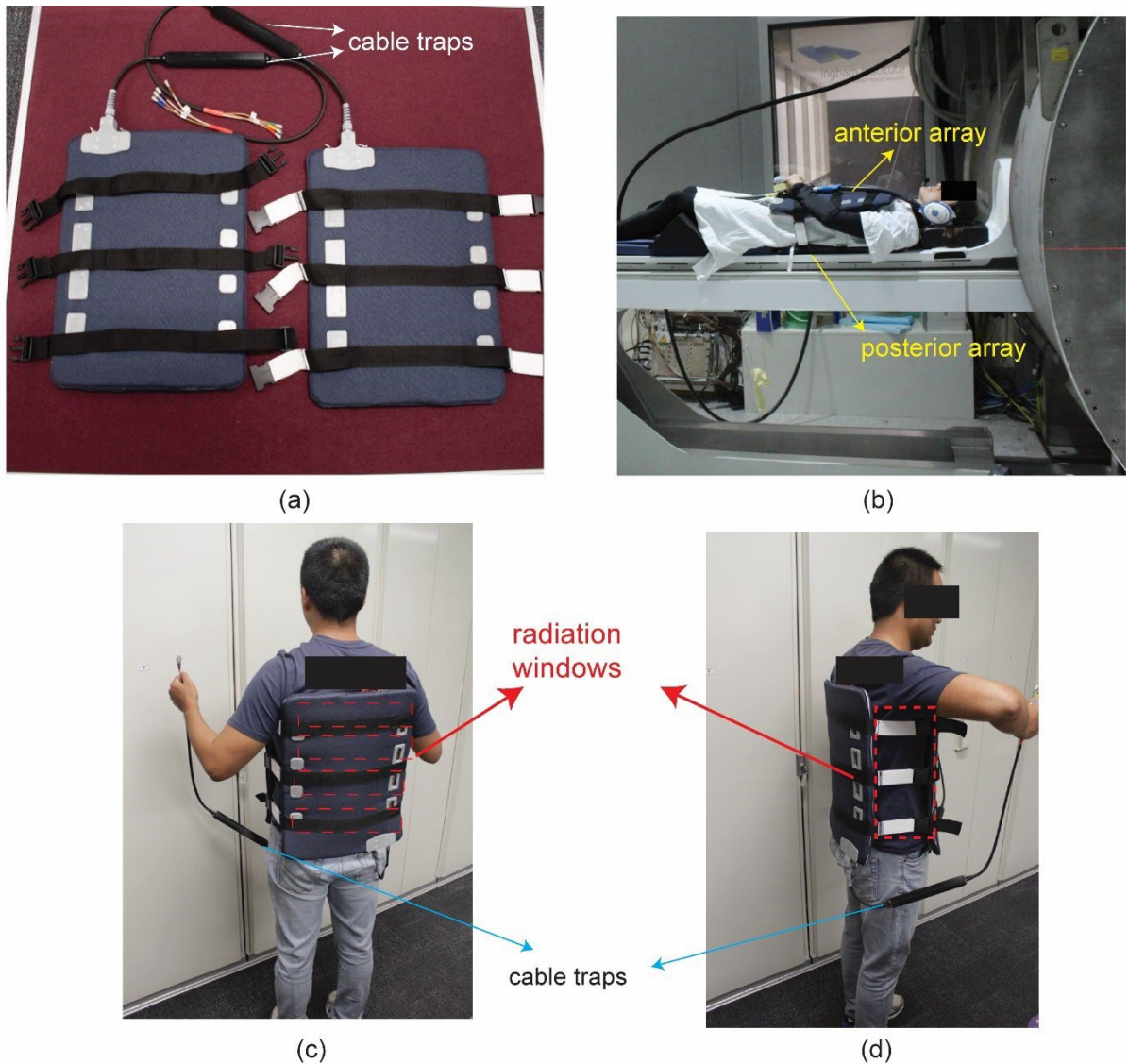

Figure 2(a) Two sub-arrays of the proposed torso coil array. (b) Coil – patient setup in the supine position on the rotating couch. (c) and (d) Coil – patient setup in standing position.

A 20 L water phantom was used for imaging and SNR comparison between the proposed coil array and the whole-body coil. The size of the phantom is 390 mm x 360 mm x 195 mm, which is similar to the size of the human mid trunk. The solution was made of 1.24 g $NiSO_4 \times 6 H_2O$ and 2.62 NaCl per 1000 mL distilled water. The *in vivo* images were acquired from a 30-year-old healthy female volunteer and a 36-year-old healthy male volunteer. All data were acquired on the 1.0 T Australian MRI-Linac system located at the Liverpool Hospital (Sydney, Australia) with the Siemens MRI console (Siemens Healthcare, Erlangen, Germany). All imaging protocols were approved by the Southwestern Sydney Local Health District Human Research Ethics Committee. Signed consent forms were obtained from volunteers for this study.

The S-parameters were measured with a four-channel Network Analyser (R&S®ZNB8, Rohde & Schwarz GmbH & Co. KG, Munich, Germany), while the torso array was fit on a volunteer at the anterior and posterior sides. To acquire S parameters for all channels, the measurements of four channels were alternated in this order: $S_{1234}$, $S_{1256}$, $S_{1278}$, $S_{3456}$, $S_{3478}$, and $S_{5678}$.

The parameters of Gradient Echo (GRE) sequence used for imaging were: TR/TE = 500/4.41 ms, image size = 256x256, FOV = 500x500 mm, bandwidth: 260 Hz/pixel, slice thickness = 5 mm. For *in vivo* imaging, Turbo Spin Echo (TSE) and TrueFISP sequences were used. The parameters of TSE sequence were: TR/TE = 505/12 ms, image size: 256x256, FOV = 400x400 mm, bandwidth: 201 Hz/pixel, slice thickness = 3 mm, echo train length = 4. The parameters of the TrueFISP sequence were: TR/TE = 3.52/1.75 ms, image size: 128x128, FOV = 400x400 mm, bandwidth: 558 Hz/pixel, slice thickness = 10 mm.

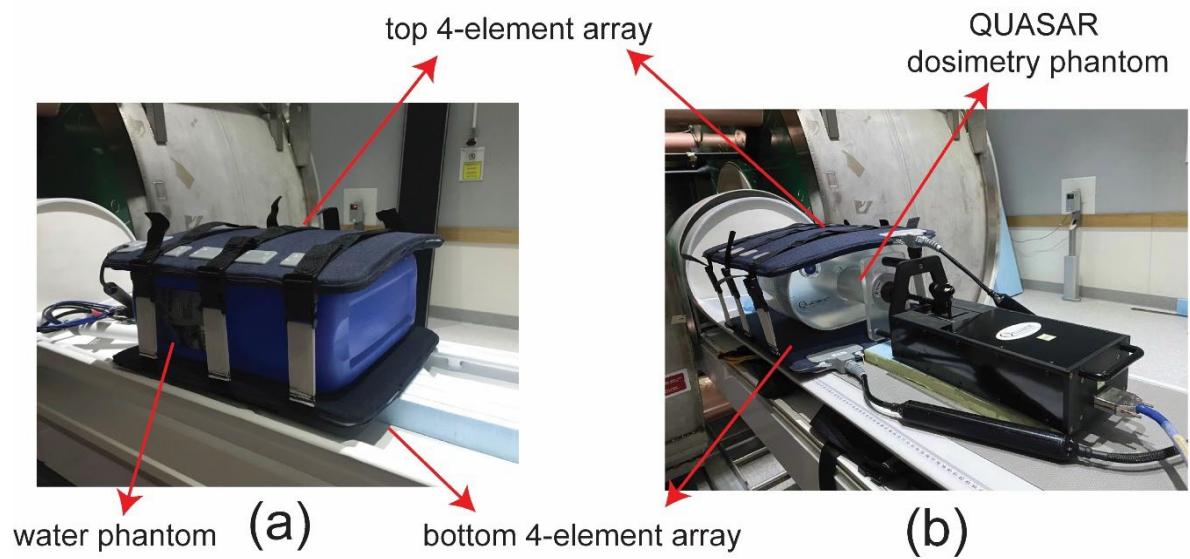

Figure 3 (a) The coil array was installed on the phantom containing homogeneous water solution for imaging and SNR comparison. (b) The coil array was installed on the QUASAR phantom for dosimetry measurements.

### 2.3 Radiolucency test

The Quasar body phantom (Modus QA, Ontario, Canada) was used for dosimetry measurement with and without the torso coil attached. To comprehensively evaluate the coil's radiolucency, the dosimetry was measured at three different treatment positions at 0º, 45º, and 90º. These angles represent the scenarios for no contact, partial contact, and maximum contact between the Linac beam and the coil. The field size of the Linac beam was set to 40 mm x 40 mm. The coil/phantom setup for these three positions is shown in Figure 3 (b).

As demonstrated in previous work[12], the background noise can be used to evaluate the interaction between the coil and the Linac beam. 120 dynamic TrueFISP images were acquired with beam on and off for every 30 images. The background noise was calculated according to the (NEMA) standard [24] by averaging the noise at four corners of the image[24].

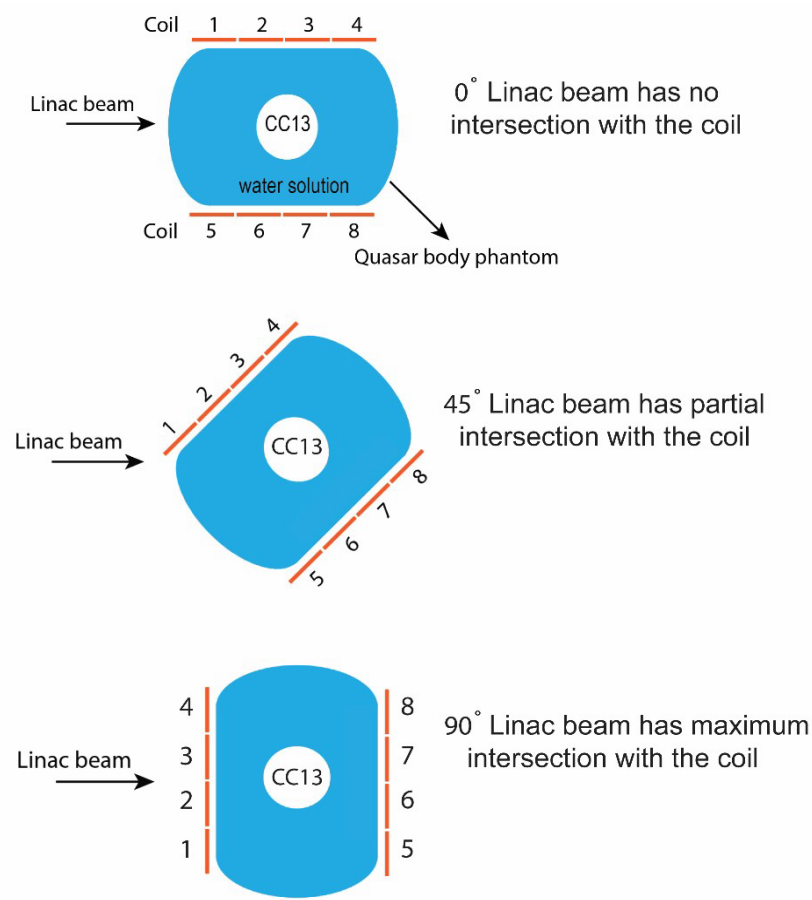

Figure 4 Dosimetry measurement setup at different treatment angles. At 0⁰, the Linac beam has zero intersection with the torso coil; at 90⁰, the intersection between the Linac beam and the coils is maximized.

## 2.4 Signal-to-noise ratio (SNR) calculation

### 2.4.1 Theoretical SNR

While comparing SNR of different coils, both transmit profile ($B_1^+$) and receive profile ($BB_1^-$) [25] need to be considered. Because the $B_1^+$ determines the strength and areas the magnetization is flipped and the $B_1^-$ denotes the signal reception capability of the coil.

The SNR is calculated according to [26] :

$$SNR = \frac{V_{signal}}{V_{noise}} \propto \frac{B_0^2 \sin(\gamma V\tau|B_1^+|)|B_1^-|}{\sqrt{P_{sample}}} \qquad (1)$$

where $P_{sample}$ denotes the power absorbed by the subject. For this work, the transmission was delivered with the same body coil; thus $B_1^+$ can be taken out of Eq. 1. Both coils used 1W power in the simulation to acquire the $B_1^-$ maps, and the SNR ratio of two coils can be written as:

$$\frac{SNR_{torso\ coil}}{SNR_{body\ coil}} \propto \frac{|B_{1\ torso\ coil}^-|}{|B_{1\ body\ coil}^-|} \qquad (2)$$

where the $B_1^-$ is calculated as [25]:

$$B_1^- = \frac{(B_x - iB_y)^*}{2} \qquad (3)$$

where $B_x$ and $B_y$ are the field vector components calculated in Sim4Life, $i$ is the imaginary unit, * asterisk indicates the complex conjugation.

### 2.4.2 SNR measurement in MRI images

The SNR calculation in the MRI images follows the National Electrical Manufacturers Association (NEMA) standard[24]. The same excitation voltage (306 V) was used for both coils with the GRE sequence. The noise images produced with 0 volts excitation voltage were acquired for noise evaluation and noise correlation matrix calculation. The background noise used for the radiolucency test with the successive 120 TrueFISP images were calculated by averaging four corners as specified in the NEMA standards.

## 3. Results

### 3.1 Bench measurement and noise correlation matrix

In order to acquire realistic measurements, both bench-measured S parameters and noise correlation matrices were obtained when the coils were worn on the volunteer (Figures 2(c) and (d)). As shown in Figure 5(a), the measured $S_{11}$ of all channels was lower than - 30dB at 42 MHz. The 8x8 measured $S_{xy}$ parameters, and the calculated noise correlation are shown in Figures 5 (b) and (c), respectively. The corresponding values (dB) are shown in Tables I and II. These results demonstrate all coils were tuned and matched well, and all channels were decoupled well.

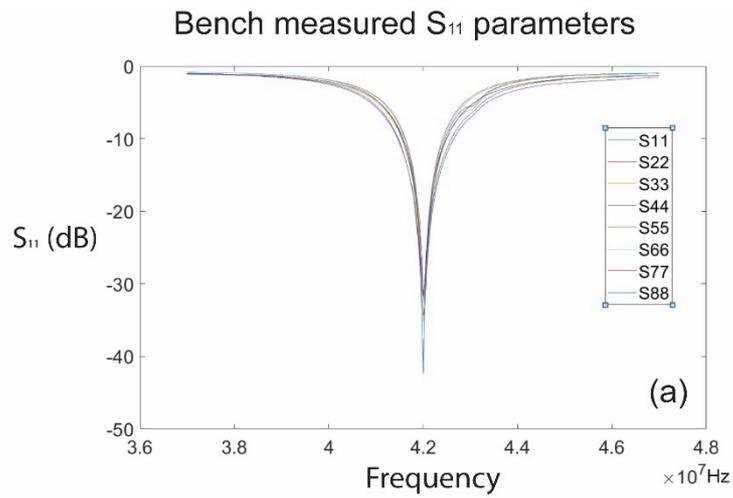

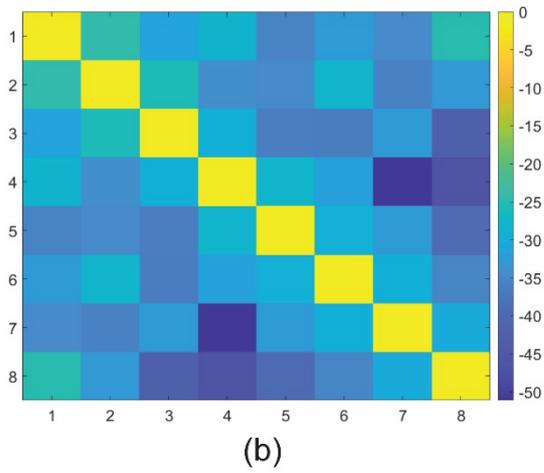
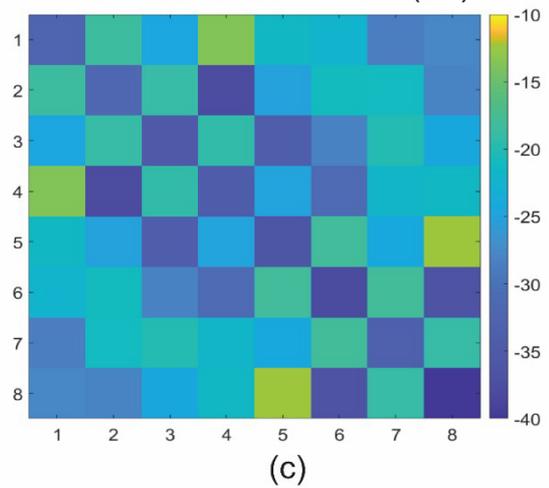

Figure 5 (a) and (b) are bench-measured $S_{11}$ and $S_{xy}$ with a four-channel Vector Network Analyser (R&S®ZNB8). (c) The noise correlation matrix was calculated from noise images acquired with 0 V on the transmit RF pulse.

Table I. Bench measured S parameters of all coils

|  | Coil 1 | Coil 2 | Coil 3 | Coil 4 | Coil 5 | Coil 6 | Coil 7 | Coil 8 |
|---|---|---|---|---|---|---|---|---|
| Coil 1 | -32.5 | -18.3 | -24.4 | -13.7 | -21.4 | -22.2 | -28.5 | -27.7 |
| Coil 2 | -18.3 | -31.7 | -19.0 | -38.0 | -25.0 | -21.0 | -21.1 | -28.1 |
| Coil 3 | -24.4 | -19.0 | -35.3 | -19.34 | -34.0 | -28.1 | -20.1 | -24.4 |
| Coil 4 | -13.7 | -38.0 | -19.3 | -34.4 | -24.9 | -31.3 | -21.9 | -21.4 |
| Coil 5 | -21.4 | -25.0 | -34.0 | -25.0 | -36.1 | -18.1 | -24.1 | -12.3 |
| Coil 6 | -22.2 | -21.0 | -28.1 | -31.3 | -18.1 | -38.2 | -18.0 | -36.9 |
| Coil 7 | -28.5 | -21.1 | -20.1 | -21.9 | -24.1 | -18.0 | -33.7 | -19.1 |
| Coil 8 | -27.7 | -28.1 | -24.4 | -21.4 | -12.3 | -36.9 | -19.1 | -42.4 |

Table II. Noise correlation matrix (dB) calculated from acquired 0-volt noise images

|  | Coil 1 | Coil 2 | Coil 3 | Coil 4 | Coil 5 | Coil 6 | Coil 7 | Coil 8 |
|---|---|---|---|---|---|---|---|---|
| Coil 1 | 0 | -24.6 | -31.4 | -28.0 | -35.7 | -32.7 | -34.7 | -24.9 |
| Coil 2 | -24.6 | 0 | -25.6 | -34.3 | -34.7 | -27.9 | -35.9 | -32.8 |
| Coil 3 | -31.4 | -25.6 | 0 | -29.5 | -36.3 | -36.5 | -32.7 | -42.8 |
| Coil 4 | -28.0 | -34.3 | -29.5 | 0 | -27.9 | -32.0 | -51.1 | -46.6 |
| Coil 5 | -35.7 | -34.7 | -36.3 | -27.9 | 0 | -29.3 | -32.6 | -39.8 |
| Coil 6 | -32.7 | -27.9 | -36.5 | -32.0 | -29.3 | 0 | -29.5 | -35.2 |
| Coil 7 | -34.7 | -35.9 | -32.7 | -51.1 | -32.6 | -29.5 | 0 | -30.3 |
| Coil 8 | -24.9 | -32.8 | -42.8 | -46.6 | -39.8 | -35.2 | -30.3 | 0 |

### 3.2 Phantom imaging

#### 3.2.1 SNR comparison

The simulated SNR maps of the whole-body coil and the 8-element torso coil array are compared in Figures 6 (a) and (b). Although the whole-body coil (b) shows a very uniform distribution, the SNR is much lower than that of the torso coil. However, the torso coil exhibits a typical field pattern caused by coil sensitivity weightings. As shown in Table III, the simulated SNR of the torso coil is three times and eight times higher than the whole-body coil at the center and surface regions, respectively.

The GRE images acquired with two coils are shown in Figures 6 (c) and (d). The coarse-grained granite texture in Figure 6(c) implies higher noise, whereas the image acquired with the torso coil in (d) shows very smooth textures. The 0 volts images for both coils were also obtained for the noise calculation. The measured noise of the torso coil is only 8.6, about one-fourth of the whole-body coil. With the same excitation voltage (306V), two coils have comparable signal intensity in the center, but in the superficial areas, the torso coil has much higher signal intensity because of the closer proximity. Although the whole-body has very similar SNR at the center region (16.4) and surface region (15.7), they are much lower than those of the torso coil, which are 67.4 (center) and 147.4 (surface), about three and eight times higher than that of the whole-body coil.

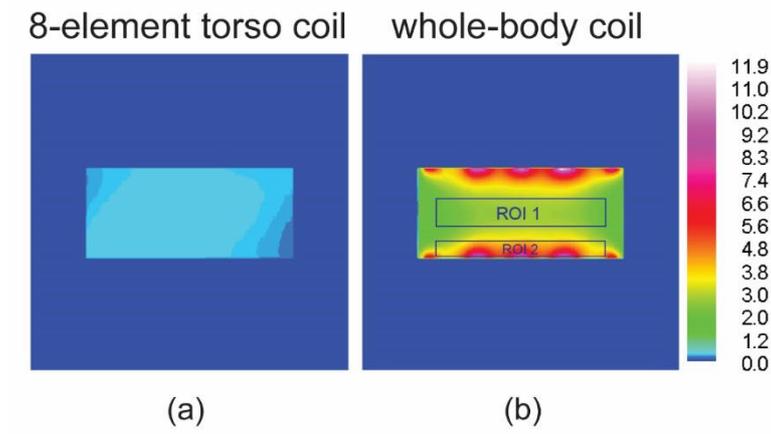

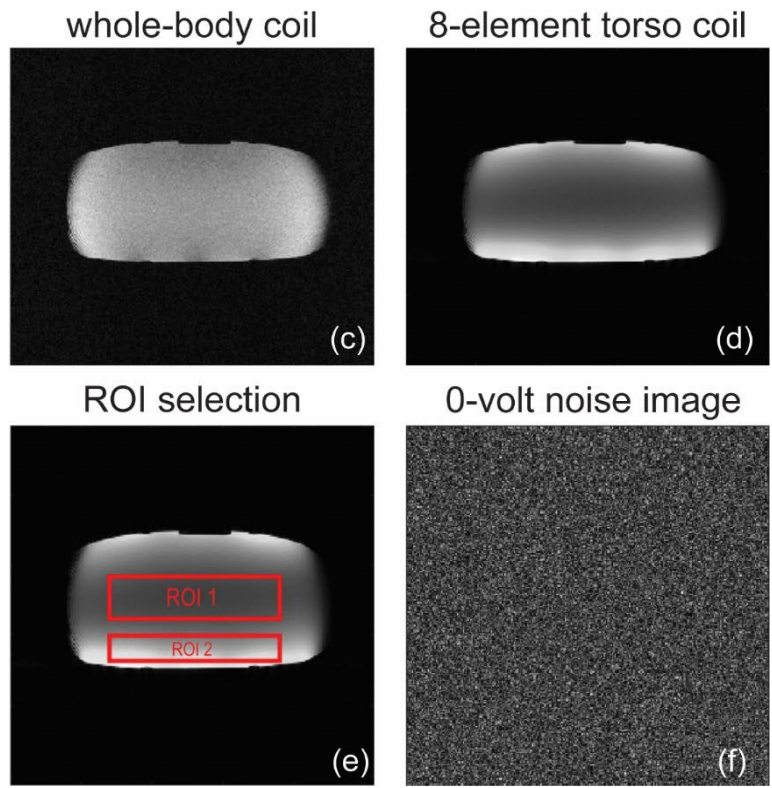

Figure 6 (a) Simulated SNR map of the whole-body coil. (b) Simulated SNR map of the 8-element torso coil array. (c) GRE image acquired with the whole-body coil when using 306V. (d) GRE image acquired with the 8-element torso coil array when using 306V. (e) ROI selection. (f) noise image when using 0 volts. ROI 1, central ROI; ROI 2, surface ROI.

Table III SNR comparison between the 8-element torso coil array and the wholebody coil

|  | 8-element torso coil array | | Whole-body coil | |
| --- | --- | --- | --- | --- |
|  | ROI 1 – Centre | ROI 2 – surface | ROI 1 – Centre | ROI 2 – surface |
| **Signal** | 505.0 | 1238.2 | 510.1 | 494.4 |
| **Noise** | 8.6 | 8.5 | 31.3 | 31.3 |
| **Measured SNR** | 58.7 | 145.7 | 16.3 | 15.8 |
| **Simulated SNR** | 67.2 | 142.4 | 16.4 | 15.7 |

### 3.2.2 Imaging at different treatment angles

Since the Linac is fixed in the Australian MRI-Linac system, the patient will need to be rotated with the rotating couch during treatment. The $B_0$ field will inevitably cross the coil, and image quality might be affected. Images were acquired at different treatment angles at 0º, 45º, 90º, 135º, 180º to evaluate the effect of the rotation. The center ROI was selected as a 15 cm$^2$ circular area with 485 pixels, and the surface ROI was selected as a 4 cm$^2$ circular area with 131 pixels. At 0º or 180º, the coils were parallel to $B_0$; thus, the signal intensity was the highest, and the coils were most efficient. When the coils rotated, they started to intersect with $B_0$, and the intersection reached the maximum at 90º. Therefore, the signal intensity at 90º was the lowest, as shown in Figure 7(b). Compared to the 0º supine position, the signal intensity at 45º had a 14% drop at the center and a 20% drop at the surface. At 90º, the signal intensity at the center dropped 33% at the center and 37% at the surface. However, since the signal reduction at the center and surface were at a similar ratio, the image quality and overall image pattern were not affected.

Table IV Signal intensity comparison at different treatment angles with the 8-element torso coil array

|  | 0º | 45º | 90º | 135º | 180º |
| --- | --- | --- | --- | --- | --- |
| Centre ROI | 331.1 | 285.2 | 223.2 | 276.7 | 313.7 |
| Surface ROI | 1246.2 | 998.4 | 786.7 | 996.1 | 1191.0 |

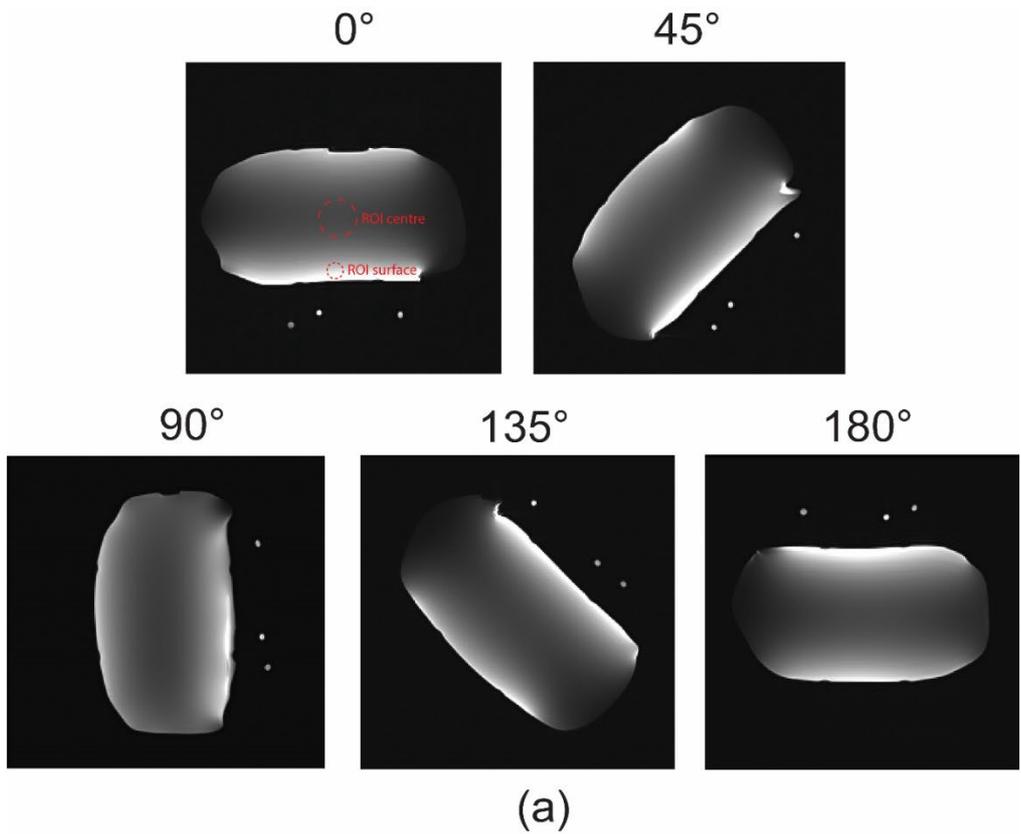
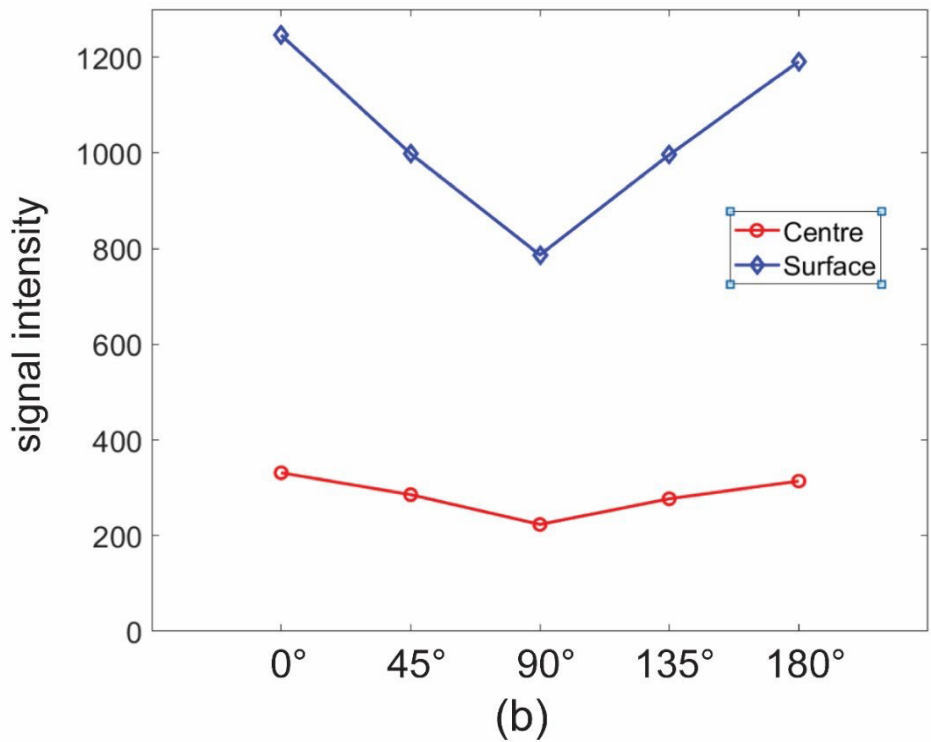

Figure 7 (a) Images acquired with the torso coil at different treatment angles. (b) Signal intensity variation at different treatment positions. Bright spots beneath the phantom are fiducial markers in the patient rotation system (PRS) of the MRI-Linac system.

## 3.3 Radiolucency validation
### 3.3.1 Dosimetry measurement with the QUASAR phantom

The torso coil array was installed on the anterior and posterior sides of the QUASAR phantom, as shown in Figure 3 (b). The dosimetry was measured at three different positions, 0º, 45º, and 90º, representing different scenarios of various intersection areas between the torso coil and the Linac beam. At 0º, the torso coil has a minimum intersection with the beam; at 90º the intersection reaches maximum. 45º is the case between the minimum and maximum intersection area. At each angle, five dosimetry measurements (Grays to far; GSF) were performed. As shown in Table V, the GSF without and with the torso coil are very similar at each angle. The GSF difference over the average GSF is under 1% for all three angles. Additionally, the GSF of five measurements within each group was also very consistent. Therefore, we can conclude the torso coil is radiolucent to the Linac beam.

Table V Dosimetry measurements with and without the presence of the torso coil

| Grays so far (GSF) Angles | Measure 1 | Measure 2 | Measure 3 | Measure 4 | Measure 5 | Mean | STD | STD/Mean |
|---|---|---|---|---|---|---|---|---|
| 0 degree – no torso coil | 33.5 | 33.2 | 33.7 | 33.6 | 33.5 | 33.5 | 0.2 | 0.6% |
| 0 degree - with torso coil | 33 | 33.3 | 33.7 | 33.3 | 33.3 | 33.3 | 0.2 | 0.6% |
| Difference (0 degree) | 0.5 | -0.1 | 0 | 0.3 | 0.2 | 0.2 | | |
| Difference/Measure (0 degree) | 1.5% | -0.3% | 0 | 0.9% | 0.6% | 0.54% | | |
| 45 degree – no torso coil | 36.5 | 37.2 | 37.0 | 36.7 | 37.2 | 36.9 | 0.3 | 0.8% |
| 45 degree - with torso coil | 36.3 | 36.5 | 36.7 | 37.0 | 36.6 | 36.6 | 0.3 | 0.8% |
| Difference (45 degree) | 0.2 | 0.7 | 0.3 | -0.3 | 0.6 | 0.3 | | |
| Difference/Measure (45 degree) | 0.6% | 1.9% | 0.8% | -0.8% | 1.6% | 0.82% | | |
| 90 degree – no torso coil | 41.2 | 41.7 | 41.4 | 41.6 | 40.9 | 41.4 | 0.4 | 1.0% |
| 90 degree - with torso coil | 40.7 | 41.2 | 41.2 | 41.2 | 41.2 | 41.2 | 0.1 | 0.2% |
| Difference (90 degree) | 0.5 | 0.5 | 0.2 | 0.4 | -0.3 | 0.26 | | |
| Difference/Measure (90 degree) | 1.2% | 1.2% | 0.5% | 1.0% | -0.7% | 0.63% | | |

### 3.3.2 Radiolucency validation with background noise

As demonstrated in previous work [12], the background noise can be used to evaluate the interaction between the coil and the Linac beam. 120 dynamic TrueFISP images were acquired with beam on and off for every 30 images. The background noise was calculated according to the NEMA standard by averaging the noise at four corners. As shown in Figure 8, the first 60 images were acquired when the multileaf collimator (MLC) aperture was closed, and the last 60 images were acquired when the MLC aperture was open for a 4x4 cm area, which is realistic for tumor treatment. The noise among all four tests was consistent, and the variation of the mean noise was less than 1%. The results indicate that the torso coil is radiolucent to the Linac beam and thus can be used during radiotherapy without compromise on performance.

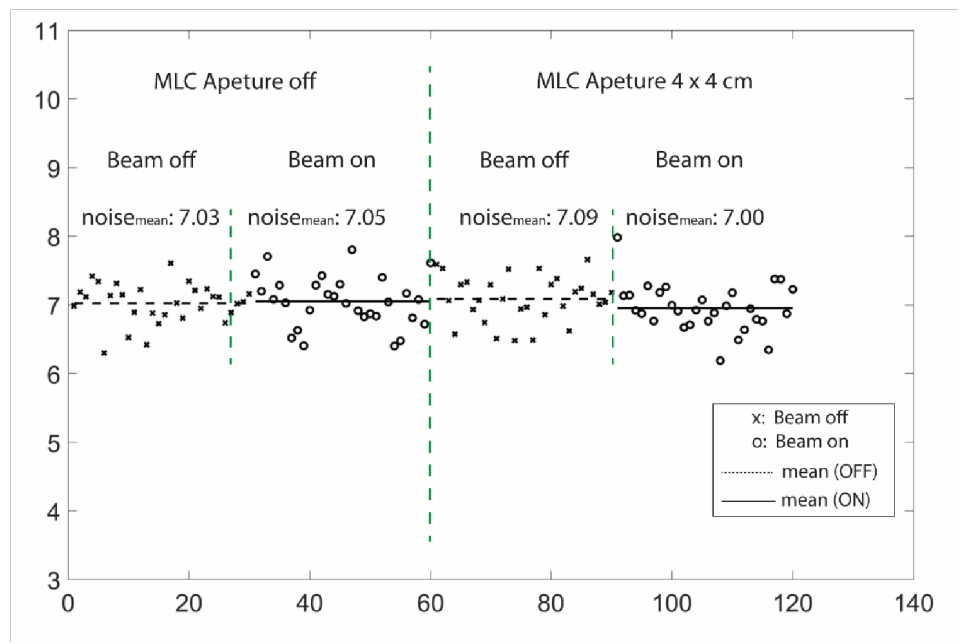

Figure 8 Background noise plot with dynamic TrueFISP sequence. The first 60 images were acquired with the MLC aperture closed, and the last 60 images were acquired with a 40 x 40 mm MLC aperture. Within each set, the first 30 images were acquired when the beam was off, and the last 30 images were acquired when the beam was on. The horizontal lines are the mean value within each interval.

### 3.4 *In vivo* imaging

The coronal abdominal images acquired with TrueFISP and TSE sequences are shown in Figure 9. The images acquired with the torso coil in (a) and (b) have much better quality than the images acquired with the whole-body coil in (c) and (d). The elevated noise in images (c) and (d) compromised their readability and diagnostic value, with less explicit boundaries and detailed structures. The torso coil is superior to the whole-body coil with the capability of exhibiting more refined structures marked with red circles and arrows in Figures 9 (a) and (b). One of the primary goals of using the simultaneous multi-element torso array is the fast-imaging capability because the MRI-Linac system needs real-time guidance for tumor tracking. As shown in (f), when the reduction factor is 2, not much difference is detected compared to the fully sampled case in (e). No severe aliasing artifacts can be found even at a reduction factor of 3, except for the slightly elevated noise.

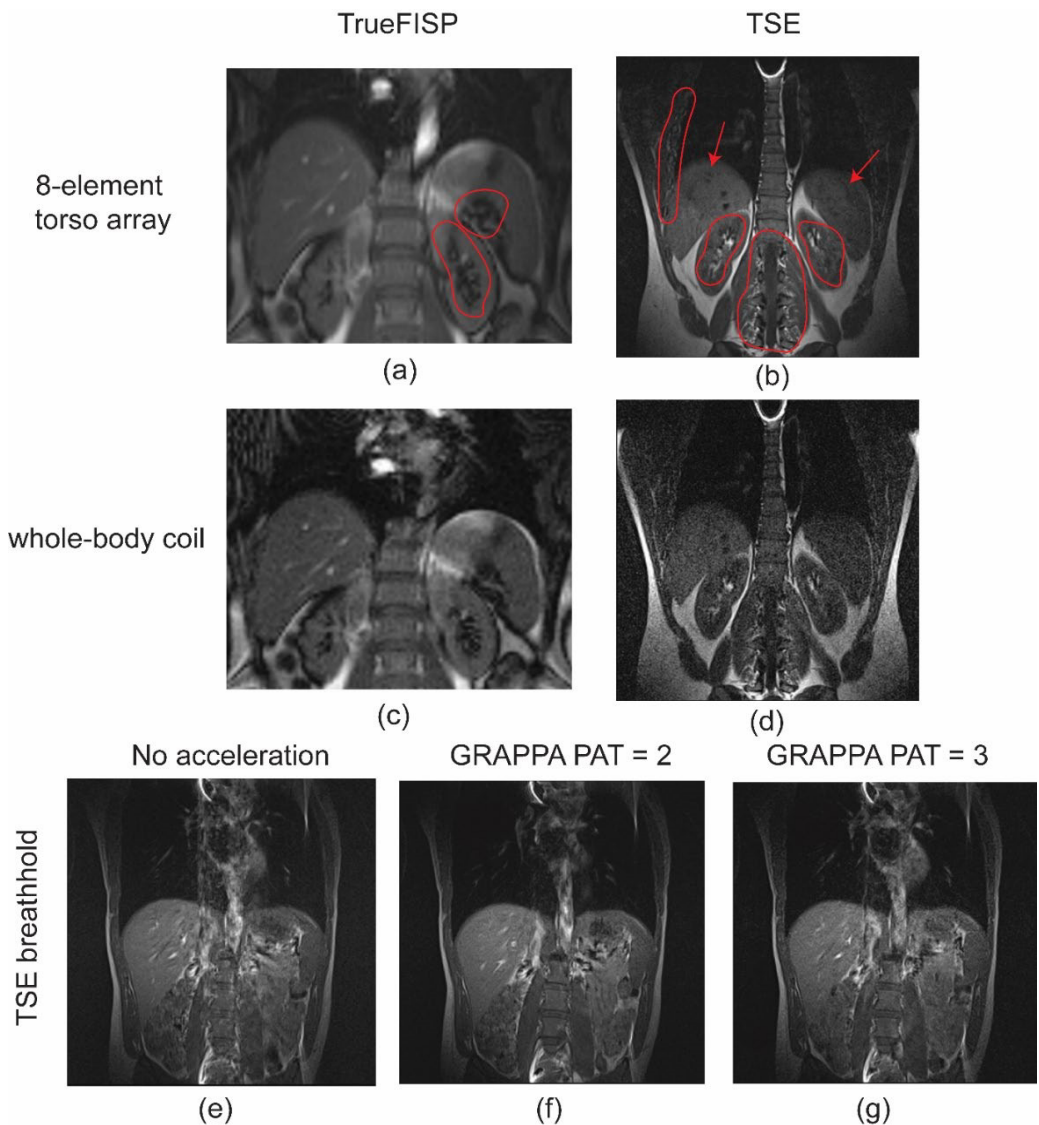

Figure 9 (a) and (b) are the images acquired with the 8-element torso array using TrueFISP and TSE sequences. (c) and (d) are images acquired with the whole-body coil using TrueFISP and TSE sequences. (e), (f) and (g) are images acquired using TSE sequence when the volunteer was holding their breath. The GRAPPA was used to accelerate imaging and the PAT factor was 1, 2 and 3, respectively. Red circles and arrows indicate the areas that have better details.

## 4. Discussion

The developed 8-element torso coil array addresses the needs of the MRI-Linac system: (1) it can provide much higher SNR than the existing whole-body coil, which can be used for better tumor characterization. (2) It can accelerate the imaging up to a factor of three without obvious aliasing artifacts, facilitating real-time tumor tracking during radiotherapy. (3) The images produced by this coil array have consistent quality regardless the treatment angles. This improves the potential accuracy of radiotherapy. However, a few points deserve discussion which could further improve the image quality or better understanding the performance of this coil array.

### 4.1 Surface sensitivity bias removal

As shown in Figures 6 and 7, sensitivity bias is often seen in images acquired using surface coils. One of the most used methods for bias removal is "prescan normalization". However, limited by the current software version on the scanner, this function is temporarily unavailable and will be implemented in the future. Alternatively, the authors have tried using the method described in [27] and the sensitivity bias have been successfully removed.

### 4.2 Number of coil elements

The current design has two sub-arrays and each sub-array has 4 coil elements. This is a balanced design with consideration of the patient rotation, coil performance, and engineering robustness. There are multiple theoretical choices when designing a body coil array. For example, each sub-array can have two rows of coils and each row could have four coils (2x4) or even 2x8 structures. These designs can surely increase the fast-imaging capability, but at the same time, smaller coils also underperform when receiving signals from deep pelvic regions, such as the prostate and ovaries. Namely, although additional small coils can increase the SNR at surface areas, the SNR in the deep regions may not be improved [21]. Small coils also have less area coverage; thus, those coils that are far away from the ROI may collect more noise than useful anatomical information. Therefore, active coil selection techniques are needed for optimal coil combination [28,29]. In addition, more complex decoupling networks could reduce the engineering and coil/beam interaction robustness.

### 4.3 Coil/beam interaction when MLC aperture was fully open

As shown in Figure 8, the background noise does not change when the beam was on/off or the aperture is on/off. However, out of interest, we also investigated the situation of a fully opened MLC aperture, although this is not a realistic setup during radiotherapy. While the aperture is fully open, the beam will inevitably interact with all coil elements and related electronics, potentially producing higher background noise. In the experiment, we acquired 60 dynamic TrueFISP images when the aperture was off and fully open. Although no artifacts are visible in the images, the background noise increased by just 14% when the MLC aperture was fully open. However, this is still much smaller than the previously developed 6-channel prototype coil array (noise elevation: 25%)[30]. Therefore, while radiation windows have been provided to allow the Linac beam pass through without any interaction with the RF coils, it can be expected that even with a fairly large beam aperture size will not dramatically increase the background noise during radiotherapy. In addition, slight noise elevation will not affect the image guidance provided by the MRI, and as such the treatment quality. More importantly, as shown in Table V, the radiotherapy dose delivered to the target area was not affected by the coil array at any of the imaging positions.

## 5. Conclusion

An 8-element torso RF receive-only coil array was designed and manufactured for the 1.0T inline Australian MRI-Linac system. This torso coil array consists of two 4-element sub-arrays that can cover the 30 cm DSV. Bench-measured S-parameters and the experimentally acquired noise correlation matrix demonstrated the coils were well decoupled and worked efficiently at the resonant frequency. The torso coil has three times higher SNR at the center and eight times higher SNR at the superficial regions than the whole-body coil. The much-improved SNR provided high-quality images with more explicit details and sharper boundaries. Higher SNR means image quality can be traded for imaging speed, with the use of higher acceleration factors. Images acquired at different rotation angles exhibit the robustness and stable performance of this coil array. The dosimetry measurements were consistent with and without the torso coil attached on the QUASAR phantom at different treatment angles, demonstrating the radiolucency of the manufactured coil array. With simultaneous multiple receive coils, this torso array could shorten the imaging duration to one third using the system built-in GRAPPA algorithm. This can facilitate real-time tumor tracking during treatment with powerful algorithms such as Compressed Sensing[16] and Neural Network based methods[31-33].